\documentstyle[epsfig,graphicx]{aipproc_old}

\begin{document}
\newcommand{\beq}{\begin{equation}}
\newcommand{\eeq}{\end{equation}}
\newcommand{\beqar}{\begin{eqnarray}}
\newcommand{\eeqar}{\end{eqnarray}}
\newcommand{\bfig}{\begin{figure}[htb]}
\newcommand{\efig}{\end{figure}}
\newcommand{\wink}{<\!\!\!\!\!\hbox{\rm )}\;}
\title{Physics of Particle Detection \footnote{ICFA Instrumentation
    School, Istanbul, Turkey, June 28 - July 10, 1999}}

\author{Claus Grupen}
\address{Department of Physics, University of Siegen \\ D-57068 Siegen,
Germany \\ e-mail: grupen@siux00.physik.uni-siegen.de }

\vspace*{-1cm}
\maketitle
\begin{abstract}
In this review the basic interaction mechanisms of charged and neutral
particles are presented. The ionization energy loss of
charged particles is fundamental to most particle detectors and is
therefore described in more detail. The production of electromagnetic
radiation in various spectral ranges leads to the detection of charged
particles in scintillation, Cherenkov and transition radiation counters.
Photons are measured via the photoelectric effect, Compton scattering or
pair production, and neutrons through their nuclear interactions.

A combination of the various detector methods helps to identify
elementary particles and nuclei. At high energies absorption techniques  in
calorimeters provide additional particle identification and an accurate
energy measurement.
\end{abstract}

\section*{Introduction}
The detection and identification of elementary particles and nuclei is
of particular importance in high energy, cosmic ray and nuclear
physics \cite{r:cgru:[1],r:cgru:[2],r:cgru:[3],r:cgru:[4],r:cgru:[5],r:cgru:[6]}.  
Identification means that the mass of the particle and
its charge is determined. In elementary particle physics most
particles have unit charge. But in the study e.g.\ of the chemical
composition of primary cosmic rays different charges must be
distinguished.

Every effect of particles or radiation can be used as a working
principle for a particle detector.

The deflection of a charged particle in a magnetic field determines
its momentum $p$; the radius of curvature $\rho$ is given by 
\beq 
\rho \propto \frac{p}{z} = \frac{\gamma m_0 \beta c}{z}
\eeq 
where $z$ is the particle's charge, $m_0$ its rest mass and
$\beta= \frac{v}{c}$ its velocity. The particle velocity can be
determined e.g.\ by a time-of-flight method yielding 
\beq 
\beta \propto \frac{1}{\tau} \quad,
\eeq 
where $\tau$ is the flight time. A calorimetric measurement provides a 
determination of the
kinetic energy 
\beq 
E^{\rm kin} = (\gamma - 1) m_0 c^2
\eeq
where $\gamma = \frac{1}{\sqrt{1-\beta^2}}$ is the Lorentz factor.

From these measurements the ratio of $m_0/z$ can be inferred, i.e.\ 
for singly charged particles we have already identified the particle.
To determine the charge one needs another $z$-sensitive effect, e.g.\ 
the ionization energy loss 
\beq 
\frac{{\rm d}E}{{\rm d}x} \propto
\frac{z^2}{\beta^2} \ln (a \beta \gamma)
\eeq
($a$ is a material dependent constant.)

Now we know $m_0$ and $z$ separately. In this way even different
isotopes of elements can be distinguished.

The basic principle of particle detection is that every physics effect
can be used as an idea to build a detector. In the following we
distinguish between the interaction of charged and neutral particles.
In most cases the observed signature of a particle is its ionization,
where the liberated charge can be collected and amplified, or its
production of electromagnetic radiation which can be converted into a
detectable signal. In this sense neutral particles are only detected
indirectly, because they must first produce in some kind of
interaction a charged particle which is then measured in the usual
way.

\section*{Interaction of Charged Particles}
\subsection*{Kinematics}
%

Four-momentum conservation allows to calculate the maximum energy
transfer of a particle of mass $m_0$ and velocity $v=\beta c$ to an
electron initially at rest to be \cite{r:cgru:[2]}

\beq
E^{\rm max}_{\rm kin} = \frac{2 m_e c^2 \beta^2 \gamma^2}{1 + 2 \gamma
\frac{m_e}{m_0} + \left( \frac{m_e}{m_0} \right)^2} = \frac{2 m_e
p^2}{m_0^2 + m_e^2 + 2m_e E/c^2} \quad,
\label{E:cgru:5}
\eeq
here $\gamma = \frac{E}{m_0 c^2}$ is the Lorentz factor, $E$ the total
energy and $p$ the momentum of the particle. \\
For low energy particles heavier than the electron ($2 \gamma
\frac{m_e}{m_0} \ll 1$; $\frac{m_e}{m_0} \ll 1$) eq. \ref{E:cgru:5} 
reduces to
\beq
E^{\rm max}_{\rm kin} = 2 m_e c^2 \beta^2 \gamma^2 \quad.
\eeq
For relativistic particles ($E_{\rm kin} \approx E$; $pc \approx E$) one gets
\beq
E^{\rm max} = \frac{E^2}{E + m_0^2 c^2/2 m_e} \quad.
\label{E:cgru:7}
\eeq
For example, in a $\mu$-$e$ collision the maximum transferable energy is
\beq
E^{\rm max} = \frac{E^2}{E+11} \hspace{2cm} E \mbox{ in GeV}
\eeq
showing that in the extreme relativistic case the complete energy can be
transferred to the electron.

If $m_0 = m_e$, eq. \ref{E:cgru:5}  is modified to
\beq
E^{\rm max}_{\rm kin} = \frac{p^2}{m_e + E/c^2} = \frac{E^2 - m_e^2 c^4}{E
+ m_e c^2} = E - m_e c^2 \quad.
\eeq

\subsection*{Scattering}
\subsubsection*{Rutherford Scattering}
\begin{figure}[ht!] 
\begin{center}
\includegraphics[bb=86 295 504
584,height=5.6cm,width=0.5\textwidth,clip]
{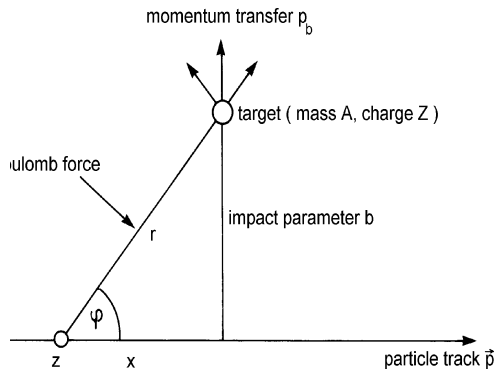}
\vspace{5pt}
\caption{Kinematics of Coulomb scattering of a particle of charge $z$ on a
target of charge $Z$}
\label{F:cgru:3}
\end{center}
\end{figure}

The scattering of a  particle of charge $z$ on a target of nuclear charge
$Z$ is mediated by
the electromagnetic interaction (figure \ref{F:cgru:3}).

The Coulomb force between the incoming particle  and the target is written
as
\beq
\vec{F} = \frac{z \cdot e \cdot Z \cdot e}{r^2} \frac{\vec{r}}{r} \quad.
\eeq
For symmetry reasons the net momentum transfer is only perpendicular to
$\vec{p}$ along the impact parameter $b$
\beq
p_b = \int^{+\infty}_{-\infty} F_b {\rm d} t = \int^{+\infty}_{-\infty} \frac{z
\cdot Z \cdot e^2}{r^2} \cdot \frac{b}{r} \cdot \frac{{\rm d}x}{\beta c} \quad,
\eeq
with $b= r \sin \varphi$, d$t = {\rm d}x/v = {\rm d}x/\beta c$, and $F_b$
force perpendicular to $p$.
\beqar
p_b & = & \frac{z \cdot Z \cdot e^2}{ \beta c} \int^{+\infty}_{-\infty}
\frac{b\,
{\rm d}x}{(\sqrt{x^2 + b^2})^3} = \frac{z \cdot Z \cdot e^2}{\beta c b }
\underbrace{\int^{+\infty}_{-\infty} \frac{{\rm d} (x/b)}{\left(\sqrt{1 + 
\left(  
\frac{x}{b} \right)^2} \right)^3}}_{=2} \\
p_b & = & \frac{ 2z \cdot Z \cdot e^2}{\beta c b} = \frac{2 r_e m_e c}{b
\beta} z \cdot Z \quad,
\label{E:cgru:13}
\eeqar
where $r_e$ is the  classical electron radius.
This consideration leads to a scattering angle
\beq
\Theta = \frac{p_b}{p} = \frac{2 z \cdot Z \cdot e^2}{\beta c b} \cdot
\frac{1}{p} \quad.
\eeq
The cross section for this process is given by the well-known Rutherford
formula
\beq
\frac{{\rm d}\sigma}{{\rm d} \Omega} = 
\frac{z^2 Z^2 r^2_e}{4} 
\left( \frac{m_ec}{\beta p} \right)^2 
\frac{1}{\sin^4 \Theta/2} \quad.
\label{E:cgru:15}
\eeq

\subsubsection*{Multiple Scattering}
From eq. \ref{E:cgru:15} one can see that the average 
scattering angle $\langle
\Theta \rangle$ is zero. To characterize the different degrees of scattering
when a particle passes through an absorber one normally uses the
so-called ``average scattering angle'' $\sqrt{\langle \Theta^2 \rangle}$.
The projected angular distribution of scattering angles in this sense leads
to an average scattering angle of \cite{r:cgru:[6]}
\beq
\sqrt{\langle \Theta^2 \rangle} = \Theta_{\rm plane} = \frac{13.6\,{\rm
MeV}}{\beta cp} z \cdot \sqrt{\frac{x}{X_0}} \left\{ 1 + 0.038 \ln \left(
\frac{x}{X_0} \right) \right\}
\eeq
with $p$ in MeV/c and $x$ the thickness of the scattering medium
measured
in radiation lengths $X_0$ (see {\bf Bremsstrahlung}). The average
scattering angle
in three dimensions is
\beq
\Theta_{\rm space} = \sqrt{2}\, \Theta_{\rm plane} = \sqrt{2}\, \Theta_0 \quad.
\eeq
The projected angular distribution of scattering angles can  approximately
be represented by a Gaussian
\beq
P (\Theta) {\rm d} \Theta = \frac{1}{\sqrt{2 \pi} \Theta_0} \exp \left\{ -
\frac{\Theta^2}{2 \Theta^2_0} \right\} {\rm d}\Theta \quad.
\eeq

\subsection*{Energy Loss of Charged Particles}
%

Charged particles interact with a medium via electromagnetic
interactions by the exchange of photons.
If the range of photons is short, the absorption of virtual photons
constituting the field of the charged particle gives rise to ionization of
the material.
If the medium is transparent Cherenkov radiation can be emitted above
a certain threshold.
But also sub-threshold emission of electromagnetic radiation can occur, 
if discontinuities of the dielectric constant of the material are
present (transition radiation) \cite{r:cgru:[7]}.
The emission of real photons by decelerating a charged particle in a
Coulomb field also constitutes an important energy loss
(bremsstrahlung). 

%
\subsection*{Ionization Energy-Loss}
\subsubsection*{Bethe-Bloch Formula}
This energy-loss mechanism represents the scattering of charged
particles off atomic electrons, e.g.
\beq
\mu^+ + \mbox{atom} \rightarrow \mu^+ + \mbox{atom}^+ + e^- \quad.
\label{E:cgru:22}
\eeq
The momentum transfer to the electron is (see eq. \ref{E:cgru:13})
\[
p_b = \frac{2 r_e m_e c}{b \beta} z \quad,
\]
and the energy transfer in the classical approximation
\beq
\varepsilon = \frac{p_b^2}{2 m_e} = \frac{2 r_e^2 m_e c^2}{b^2 \beta^2} z^2
\quad.
\label{E:cgru:23}
\eeq
The interaction probability per (g/cm$^2$), given the atomic cross-section
$\sigma$, is
\beq
\phi ({\rm g}^{-1} {\rm cm}^2) = \frac{N}{A} \sigma [{\rm cm}^2/{\rm atom}]
\eeq
where $N$ is Avogadro's constant.

\begin{figure}[ht!] 
\begin{center}
\includegraphics[height=4.5cm,width=0.5\textwidth]{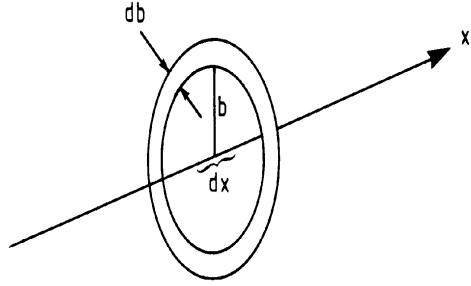}
\vspace{5pt}
\caption{Sketch explaining the differential collision probability}
\label{F:cgru:8}
\end{center}
\end{figure}

The differential probability to hit an electron in the area of an annulus with
radii $b$ and $b+{\rm d}b$
(see figure \ref{F:cgru:8}) with an energy transfer between $\varepsilon$ and
$\varepsilon + {\rm d} \varepsilon$ is
\beq
\phi (\varepsilon) {\rm d} \varepsilon = \frac{N}{A} 2 \pi b\,{\rm d}b\,Z 
\quad,
\label{E:cgru:25}
\eeq
because there are $Z$ electrons per target atom.

\begin{figure}[ht!] 
\begin{center}
\includegraphics[angle=0.2,height=7.5cm,width=0.55\textwidth,clip]{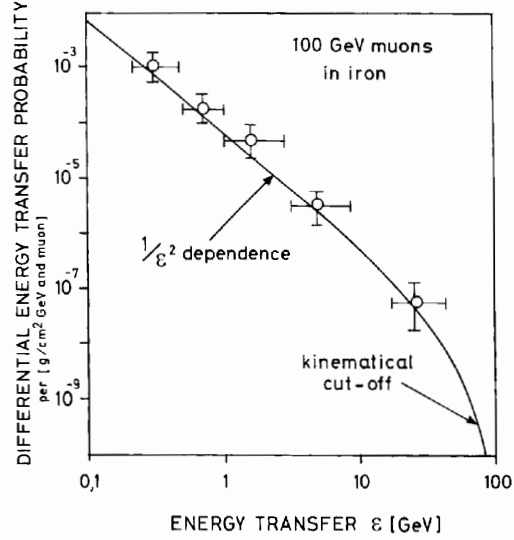}
\vspace{5pt}
\caption{$1/\varepsilon^2$-dependence of the knock-on electron
production probability \protect\cite{r:cgru:[8]} }
\label{F:cgru:9}
\end{center}
\end{figure}

Inserting $b$ from eq. \ref{E:cgru:23} into eq. \ref{E:cgru:25} gives
\beqar
b^2 & = & \frac{2 r_e^2 m_e c^2}{\beta^2} z^2 \cdot \frac{1}{\varepsilon}
\nonumber \\
2|b\,{\rm d}b| & = & \frac{2 r_e^2 m_e c^2}{\beta^2} z^2 \cdot \frac{{\rm
d}\varepsilon}{\varepsilon^2} \nonumber \\
\phi (\varepsilon) {\rm d} \varepsilon & = & \frac{N}{A} \pi \frac{2 r_e^2 m_e
c^2}{\beta^2} z^2 \cdot Z \cdot \frac{{\rm d}\varepsilon}{\varepsilon^2}
\nonumber \\
& = & \frac{2 \pi r_e^2 m_e c^2 N}{\beta^2} \cdot \frac{Z}{A} \cdot z^2 \cdot
\frac{{\rm d}\varepsilon}{\varepsilon^2} \quad,
\label{E:cgru:26}
\eeqar
showing that the energy spectrum of $\delta$-electrons or knock-on
electrons follows an $1/\varepsilon^2$ dependence (figure \ref{F:cgru:9}, 
\cite{r:cgru:[8]}).

The energy loss is now computed from eq. \ref{E:cgru:25} by integrating over 
all possible impact parameters \cite{r:cgru:[5]}
\beqar
- {\rm d} E & = & \int^{\infty}_0 \phi (\varepsilon) \cdot \varepsilon \cdot {\
d}x  \nonumber \\
& = & \int^{\infty}_0 \frac{N}{A} 2 \pi b \cdot {\rm d} b \cdot Z \cdot 
\varepsilon \cdot
{\rm d}x \nonumber \\
-\frac{{\rm d}E}{{\rm d}x} & = & \frac{2\pi N}{A} \cdot Z \int^{\infty}_0
\varepsilon \cdot b \cdot {\rm d} b \nonumber \\
& = & 2 \pi \frac{Z \cdot N}{A} \cdot \frac{2 r_e^2 m_e c^2}{\beta^2} z^2
\int^{\infty}_0 \frac{{\rm d} b}{b} \quad.
\label{E:cgru:27}
\eeqar
This classical calculation yields an integral which diverges for $b=0$ as
well as for $b=\infty$. This is not a surprise because one would not expect
that our approximations hold for these extremes.
\begin{itemize}
\item[a)] The $b=0$ case: Let us approximate the ``size'' of the
  target electron seen from the rest frame of the incident particle by
half the de Broglie wavelength. This gives a minimum impact parameter of
\beq
b_{\min} = \frac{h}{2 p} = \frac{h}{2 \gamma m_e \beta c} \quad.
\label{E:cgru:28}
\eeq
\item[b)] The $b=\infty$ case: If the revolution time $\tau_R$ of the electron
in the target atom becomes smaller than the interaction time $\tau_i$, the
incident particle ``sees'' a more or less neutral atom
\beq
\tau_i = \frac{b_{\max}}{v} \sqrt{1-\beta^2} \quad.
\eeq
\end{itemize}
The factor $\sqrt{1-\beta^2}$ takes into account that the field at high
velocities is Lorentz-contracted. Hence the interaction time is shorter. For
the revolution time we have
\beq
\tau_R = \frac{1}{\nu_Z \cdot Z} = \frac{h}{I} \quad,
\eeq
where $I$ is the mean excitation energy of the target material, which
can be approximated by
\beq
I = 10 eV \cdot Z
\eeq
for elements heavier than sulphur.

The condition to see the target as neutral now leads to
\[
\tau_R = \tau_i \qquad \Rightarrow \qquad \frac{b_{\max}}{v}
\sqrt{1-\beta^2} = \frac{h}{I}
\]
\beq
b_{\max} = \frac{\gamma h \beta c}{I} \quad.
\label{E:cgru:31}
\eeq
With the help of eq. \ref{E:cgru:28} and \ref{E:cgru:31} we can solve 
the integral in eq. \ref{E:cgru:27}
\beq
-\frac{{\rm d}E}{{\rm d}x} = 2 \pi \cdot \frac{Z}{A} N \cdot \frac{2 r_e^2 m_e
c^2}{\beta^2} z^2 \cdot \ln \frac{2 \gamma^2 \beta^2 m_e c^2}{I} \quad.
\eeq
Since for long-distance interactions the Coulomb field is screened by the
intervening  matter one has
\beq
-\frac{{\rm d}E}{{\rm d}x} = \kappa z^2 \cdot \frac{Z}{A} \frac{1}{\beta^2}
\left[ \ln \frac{2 \gamma^2 \beta^2 m_e c^2}{I} - \eta \right] \quad,
\label{E:cgru:33}
\eeq
where $\eta$ is a screening parameter (density parameter) and
\[
\kappa = 4 \pi N r^2_e m_e c^2 \quad.
\]
The exact treatment of the ionization energy loss of heavy particles leads
to \cite{r:cgru:[6]}
\beq
-\frac{{\rm d}E}{{\rm d}x} = \kappa z^2 \cdot \frac{Z}{A} \cdot
\frac{1}{\beta^2} \left[ \frac{1}{2} \ln \frac{2 m_e c^2 \gamma^2
\beta^2}{I^2} E_{\rm kin}^{\max} - \beta^2 - \frac{\delta}{2} \right]
\eeq
which reduces to eq. \ref{E:cgru:33} for $\gamma m_e/m_0 \ll 1$ and $\beta^2 -
\frac{\delta}{2} = \eta$.

\begin{figure}[ht!] 
\begin{center}
\includegraphics[angle=0.2,width=0.7\textwidth,clip]
{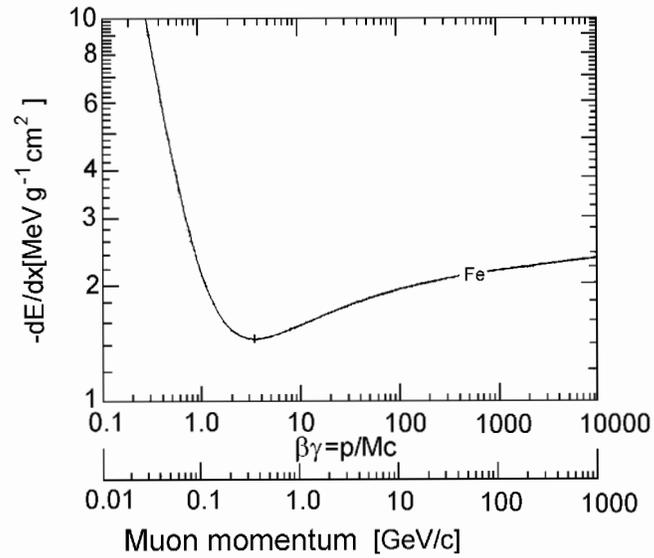}
\caption{Energy loss of muons in iron \protect\cite{r:cgru:[6]}}
\label{F:cgru:10}
\end{center}
\end{figure}

The energy-loss rate of muons in iron is shown in Figure
\ref{F:cgru:10}\cite{r:cgru:[6]}. It exhibits a
$\frac{1}{\beta^2}$-decrease until a minimum of ionization is obtained 
for $3 \leq \beta \gamma \leq 4$.

Due to the $\ln
\gamma$-term the energy loss increases again (relativistic rise, logarithmic
rise) until a plateau is reached (density effect, Fermi plateau).

The energy loss is usually expressed in terms of the area density
${\rm d}s = \rho {\rm d}x$
with $\rho$-density of the absorber.
It varies with the target material like $Z/A$ ($\leq 0.5$ for most
elements). Minimum ionizing particles lose 1.94 MeV/(g/cm$^2$) in
helium decreasing to 1.08 MeV/(g/cm$^2$) in uranium. The energy loss
of minimum ionizing particles in hydrogen is exceptionally large,
because here $Z/A = 1$.

The relativistic rise saturates at high energies because the medium
becomes polarized, effectively reducing the influence of distant
collisions. The density correction $\delta/2$ can be described by
\beq
\frac{\delta}{2} = \ln \frac{\hbar \omega_p}{I} + \ln \beta \gamma -
\frac{1}{2}
\label{E:cgru:new33}
\eeq
where
\beq
\hbar \omega_p = \sqrt{4 \pi N_e r^3_e} m_e c^2 / \alpha
\label{E:cgru:new34}
\eeq
is the plasma energy and $N_e$ the electron density of the absorbing
material. 

\begin{figure}[ht!] 
\begin{center}
\includegraphics[width=0.7\textwidth,clip]{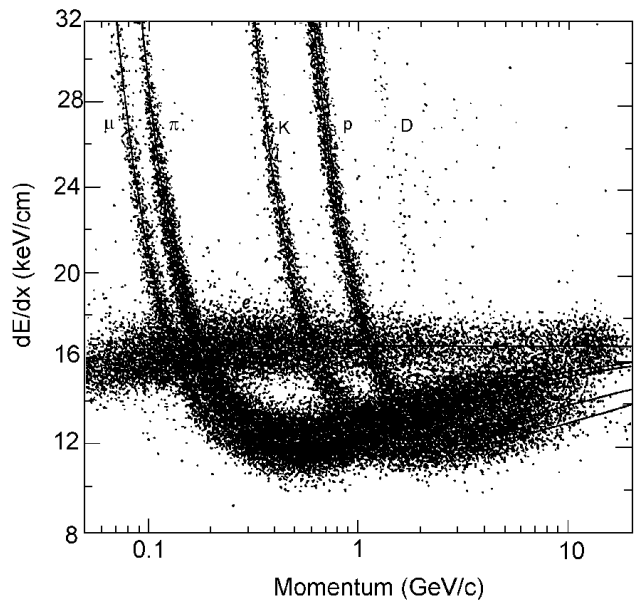}
\vspace{5pt}
\caption{Measured ionization energy loss of electrons, muons, pions, kaons,
protons and deuterons in the PEP4/9-TPC \protect\cite{r:cgru:[newNM]}}
\label{F:cgru:11}
\end{center}
\end{figure}

For gases the Fermi-plateau, which saturates the relativistic rise, is
about $60\%$ higher compared to the minimum of ionization.
Figure \ref{F:cgru:11} shows the measured energy-loss rates of 
electrons, muons, pions, kaons, protons and deuterons in the PEP4/9-TPC (185
$dE/dx$ measurements at 8.5 atm in Ar-CH$_4$ = 80 : 20)
\cite{r:cgru:[newNM]}.

\subsubsection*{Landau Distributions}
The Bethe-Bloch formula describes the average energy loss of charged
particles. The fluctuation of the energy loss around the mean is described
by an asymmetric  distribution, the Landau distribution 
\cite{r:cgru:[10],r:cgru:[11]}.

The probability $\phi (\varepsilon) {\rm d}\varepsilon$ that a singly charged
particle loses an energy between $\varepsilon$ and $\varepsilon + {\rm
d}\varepsilon$ per unit length of an absorber was (eq. \ref{E:cgru:26})
\beq
\phi (\varepsilon) = \frac{2 \pi N e^4}{m_e v^2} \frac{Z}{A} \cdot
\frac{1}{\varepsilon^2} \quad.
\label{E:cgru:35}
\eeq
Let us define

\beq
\xi = \frac{2\pi Ne^4}{m_ev^2} \cdot \frac{Z}{A} x \quad,
\eeq
where $x$ is the area density of the absorber:

\beq
\phi (\varepsilon) = \xi (x) \frac{1}{x \varepsilon^2} \quad.
\eeq
Numerically one can write

\beq
\xi = \frac{0.1536}{\beta^2} \frac{Z}{A} \cdot x \quad \mbox{[keV]} \quad,
\eeq
where $x$ is measured in mg/cm$^2$.

For an absorber of 1\,cm Ar we have for $\beta=1$
\[
\xi = 0.123\,{\rm keV} \quad.
\]
We define now
\beq
f(x, \Delta) = \frac{1}{\xi} \omega (\lambda)
\eeq
as the probability that the particle loses an energy $\Delta$ on traversing
an absorber of thickness $x$. $\lambda$ is defined to be the normalized
deviation from the most
probable energy loss $\Delta^{\rm m.p.}$
\beq
\lambda = \frac{\Delta - \Delta^{\rm m.p.}}{\xi} \quad.
\eeq
The most probable energy loss is calculated to be 
\cite{r:cgru:[10],r:cgru:[12]}
\beq
\Delta^{\rm m.p.} = \xi \left\{ \ln \frac{2 m_e c^2 \beta^2 \gamma^2 \xi}
{I^2} -
\beta^2 + 1- \gamma_E \right\} \quad,
\eeq
where $\gamma_E = 0.577 \ldots$ is Euler's constant.

\begin{figure}[!ht]
  \begin{center}
    \leavevmode
    \includegraphics[width=0.7\textwidth,clip]{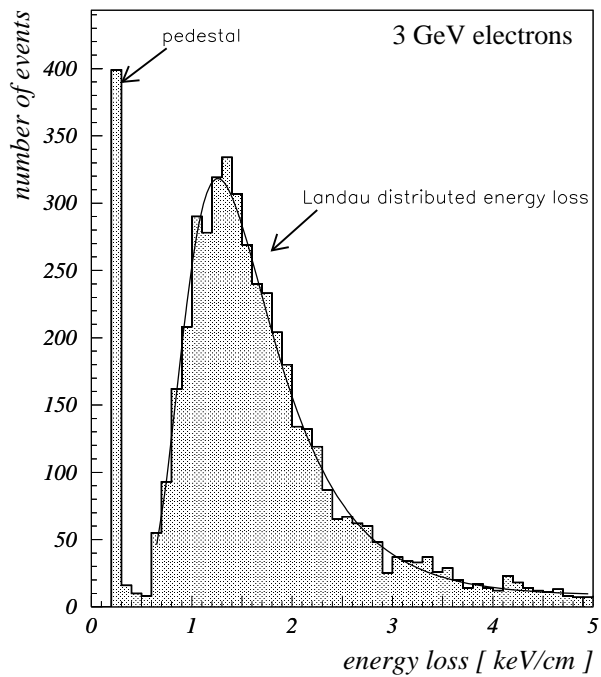}
    \vspace{5pt}
    \caption{Energy-loss distribution of 3\,GeV electrons in a thin-gap
      multiwire drift chamber \protect\cite{r:cgru:[13]}}
    \label{F:cgru:13}
    \vspace{4.7mm}
  \end{center}
\end{figure}

Landau's treatment of $f(x, \Delta)$ yields
\beq
\omega (\lambda) = \frac{1}{\pi} \int^{\infty}_0 {\rm e}^{-u \ln u - \lambda u}
\sin \pi u {\rm d} u \quad,
\eeq
which can be approximated by \cite{r:cgru:[12]}
\beq
\Omega (\lambda) = \frac{1}{\sqrt{2 \pi}} \exp \left\{ - \frac{1}{2} (\lambda +
{\rm e}^{-\lambda}) \right\} \quad.
\eeq
Figure \ref{F:cgru:13} shows the energy loss distribution of 3\,GeV 
electrons in an
Ar/CH$_4$ (80:20) filled drift chamber of 0.5\,cm thickness 
\protect\cite{r:cgru:[13]}.
According to equation \ref{E:cgru:35} the $\delta$-ray contribution to the energy 
loss falls inversely proportional to the energy transfer squared, 
producing a long tail, called Landau tail, in the energy-loss distribution 
up to the kinematical limit (see also figure \ref{F:cgru:9}).

The asymmetric property of the energy-loss distribution becomes obvious
for thin absorbers. For larger
absorber thicknesses or truncation techniques applied to thin absorbers
the Landau  distribution  gets more symmetric.

\subsection*{Scintillation in Materials}
Scintillator materials can be inorganic  crystals, organic liquids or
plastics 
and gases. The scintillation mechanism in organic crystals is an effect of
the
lattice. Incident particles can transfer energy to the lattice by creating
electron-hole pairs or taking electrons to higher energy levels below the
conduction band. Recombination of electron-hole pairs may lead to the
emission of light. Also electron-hole bound states (excitons) moving
through the lattice can emit light when hitting an activator center and
transferring their binding energy to activator levels, which subsequently
deexcite. In thallium doped NaI-crystals about 25\,eV are required to
produce one scintillation photon. The decay time in inorganic scintillators
can be quite long (1$\mu$s  in CsI (Tl); 0.62\,$\mu$s in BaF$_2$).

In organic substances the scintillation mechanism is different. Certain
types of molecules will release a small fraction ($\approx$ 3\%) of the
absorbed
energy as optical photons. This process is expecially marked in organic
substances which contain aromatic rings, such as polystyrene,
polyvinyltoluene, and naphtalene. Liquids which scintillate include toluene
or xylene \cite{r:cgru:[6]}.

This primary scintillation light is preferentially emitted in the
UV-range. The absorption length for UV-photons in the scintillation material
is rather short: the scintillator is not transparent for its own scintillation
light. Therefore, this light is transferred to a wavelength shifter which
absorbs the UV-light and reemits it at longer wavelengths (e.g.\ in the
green). Due to the lower concentration of the wavelength shifter material
the reemitted light can get out of the  scintillator and be detected by a
photosensitive device. The technique of wavelength shifting is also used
to match the emitted light to the spectral sensitivity of the photomultiplier.
For plastic scintillators the primary scintillator and wavelength shifter are
mixed with an organic material to form a polymerizing structure. In liquid
scintillators the two active components are mixed with an organic base 
\cite{r:cgru:[2]}.

About 100\,eV are required to produce one photon in an organic
scintillator. The decay time of the light signal in plastic scintillators is
substantially shorter compared to inorganic substances (e.g.\ 30\,ns in
naphtalene).

Because of the low light absorption in gases there is no need for
wavelength shifting in gas scintillators.

Plastic scintillators do not respond linearly to the energy-loss density. The
number of photons produced by charged particles is described by Birk's
semi-empirical formula \cite{r:cgru:[6],r:cgru:[17],r:cgru:[18]}
\beq
N = N_0 \frac{{\rm d}E/{\rm d}x}{1 + k_B\,{\rm d}E/{\rm d}x} \quad,
\label{E:cgru:49}
\eeq
where $N_0$ is  the photon yield at low specific ionization density, and
$k_B$ is Birk's density parameter. For 100\,MeV protons in plastic
scintillators one has d$E/$d$x \approx 10$\,MeV/(g/cm$^2$)  and $k_B
\approx 5$\,mg/(cm$^2$MeV), yielding a saturation effect of $\sim 5$\% 
\cite{r:cgru:[4]}.

For low energy losses eq. \ref{E:cgru:49} leads to a linear dependence
\beq
N = N_0 \cdot {\rm d}E/{\rm d}x \quad,
\eeq
while for very high d$E/$d$x$ saturation occurs at
\beq
N = N_0/k_B \quad.
\eeq
There exists a correlation between the energy loss of a particle that goes
into the creation of electron-ion pairs or the production of scintillation 
light because electron-ion pairs can recombine thus reducing the
d$E/$d$x|_{\rm ion}$-signal. On the other hand the scintillation light signal
is enhanced because recombination frequently leads to excited states
which deexcite yielding scintillation light.

\subsection*{Cherenkov Radiation}
A charged particle traversing a medium with refractive index $n$ with a
velocity $v$ exceeding the velocity of light $c/n$ in that medium, emits
Cherenkov radiation. The threshold condition is given by 
\beq
\beta_{\rm thres} = \frac{v_{\rm thres}}{c} \geq \frac{1}{n} \quad.
\eeq
The angle of emission increases with the velocity reaching a maximum
value for $\beta=1$, namely
\beq
\Theta_c^{\max} =  \arccos \frac{1}{n} \quad.
\eeq
The threshold velocity translates into a threshold energy
\beq
E_{\rm thres} = \gamma_{\rm thres} m_0 c^2
\eeq
yielding
\beq
\gamma_{\rm thres} = \frac{1}{\sqrt{1-\beta^2_{\rm thres}}} =
\frac{n}{\sqrt{n^2-1}} \quad.
\eeq

\noindent
The number of Cherenkov photons emitted per unit path length d$x$ is
\beq
\frac{{\rm d}N}{{\rm d}x} = 2 \pi \alpha z^2 \int \left( 1 - 
\frac{1}{n^2\beta^2}\right) \frac{{\rm d}\lambda}{\lambda^2}
\label{E:cgru:56}
\eeq
for $n(\lambda) >1$, $z$ -- electric charge of the incident particle,
$\lambda$ -- wavelength, and $\alpha$ -- fine structure constant. The  yield
of Cherenkov radiation photons is proportional to $1/\lambda^2$, but only for 
those wavelengths where the refractive index is larger than unity. Since
$n(\lambda) \approx 1$ in the X-ray region, there is no X-ray Cherenkov
emission. Integrating eq. \ref{E:cgru:56}  over the visible spectrum 
($\lambda_1 =
400$\,nm, $\lambda_2 = 700$\,nm) gives

\beqar
\frac{{\rm d}N}{{\rm d}x} & = & 2 \pi \alpha z^2 \frac{\lambda_2 -
\lambda_1}{\lambda_1 \lambda_2} \sin^2 \Theta_c \nonumber \\
& = & 490 \cdot z^2 \cdot \sin^2 \Theta_c \,[{\rm cm}^{-1}] \quad.
\eeqar

\begin{figure}[ht!]
  \begin{center}
    \includegraphics[width=0.7\textwidth,clip]{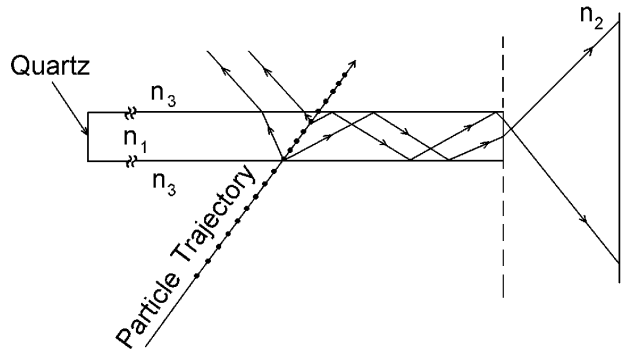}
    \vspace*{0.5cm}
    \caption{Imaging principle of a DIRC-counter
      \protect\cite{r:cgru:Adam}}
    \label{F:cgru:new7}
  \end{center}
\end{figure}

The Cherenkov
effect can be used to identify particles of fixed momentum by means of
threshold Cherenkov counters.
More information can be obtained, if the Cherenkov angle is measured
by DIRC-counters (Detection of Internally Reflected Cherenkov
light). In these devices some fraction of the Cherenkov light produced 
by a charged particle is kept inside the radiator by total internal
reflection. The direction of the photons remains unchanged and the
Cherenkov angle is conserved during the transport. When exiting the
radiator the photons produce a Cherenkov ring on a planar detector
(figure \ref{F:cgru:new7}).

\begin{figure}[ht!]
  \begin{center}
    \includegraphics[width=0.7\textwidth,clip]{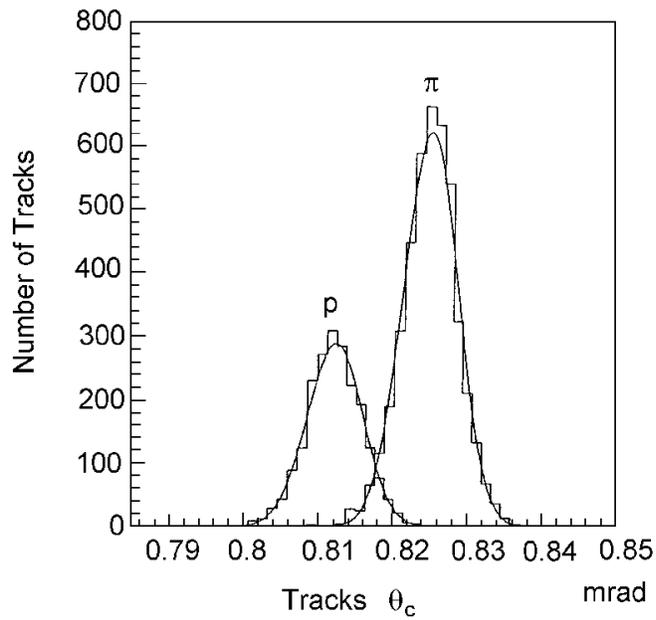}
    \caption{Cherenkov-angle distribution for pions and protons of 5.4 
      GeV/c in a DIRC-counter \protect\cite{r:cgru:Adam}.}
    \label{F:cgru:new8}
  \end{center}
\end{figure}

The pion/proton separation achieved with such a system is shown in
figure \ref{F:cgru:new8}\cite{r:cgru:Adam}.

\begin{figure}[ht!]
  \begin{center}
    \includegraphics[width=0.7\textwidth,clip]{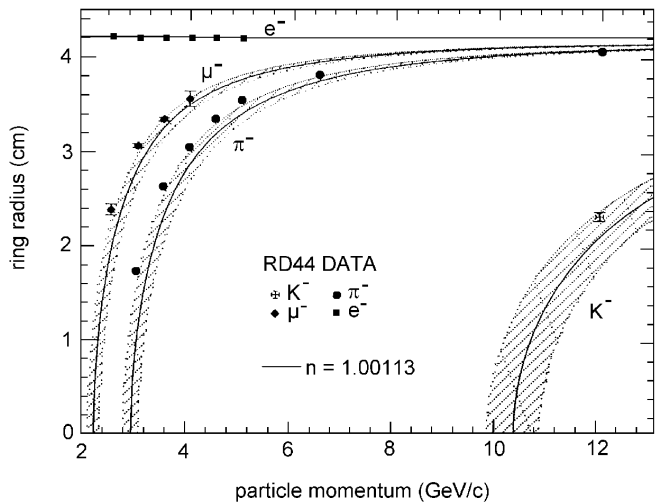}
    \caption{Cherenkov-ring radii of $e, \mu, \pi, K$ in a
      C$_4$F$_{10}$-Ar (75:25) RICH-counter. The solid curves show the
      expected radii for an index of refraction of n = 1.00113. The
      shaded regions represent a 5\% uncertainty in the absolute
      momentum scale \protect\cite{r:cgru:Debbe}}
    \label{F:cgru:new9}
  \end{center}
\end{figure}

Ring-imaging Cherenkov-counters (RICH-counters) have become
extraordinary useful in the field of elementary particles and
astrophysics. Figure \ref{F:cgru:new9} shows the Cherenkov ring radii of
electrons, muons, pions and kaons in a C$_4$F$_{10}$-Ar (75:25) filled
Rich-counter read out by a 100-channel photomultipiler of $10*10$
cm$^2$ active area \cite{r:cgru:Debbe}.

\newpage
\subsection*{Transition Radiation}
Transition radiation is emitted when a charged particle traverses a medium
with discontinuous dielectric  constant. A charged particle moving  towards
a boundary, where the dielectric constant changes, can be considered to
form together with
its mirror  charge an electric  dipole whose field strength varies in time. The
time dependent dipole field causes the emission of electromagnetic
radiation. This emission can be understood in such a way that although the
dielectric displacement $\vec{D} = \varepsilon \varepsilon_0 \vec{E}$ varies
continuously in passing through a boundary, the electric field does not.

The energy radiated from a single boundary (transition from vacuum to a
medium with dielectric constant $\varepsilon$) is proportional to the
Lorentz-factor of the incident charged particle 
\cite{r:cgru:[6],r:cgru:[17],r:cgru:[21]}:
\beq
S = \frac{1}{3} \alpha z^2 \hbar \omega_p \gamma \quad,
\eeq
where
$\hbar \omega_p$ is the plasma energy (see equation
\ref{E:cgru:new34}). For commonly used plastic radiators (styrene or similar
materials) one has

\beq
\hbar \omega_p \approx 20\,{\rm eV} \quad.
\eeq

The typical emission angle of transition radiation is proportional to
$1/\gamma$. The radiation yield drops sharply for frequencies
\beq
\omega > \gamma \omega_p \quad.
\eeq
The $\gamma$-dependence of the emitted energy originates mainly from
the hardening of the spectrum rather than from the increased photon
yield. Since the radiated photons also have energies proportional to the
Lorentz
factor of the incident particle, the number of emitted transition radiation
photons is
\beq
N \propto \alpha z^2 \quad.
\eeq
The number of emitted photons can be increased by using many transitions
(stack of foils, or foam). 
At each interface the emission probability for an X-ray photon is of
the order of $\alpha = 1/137$.
However, the foils or foams have to be of low
$Z$ material to avoid absorption in the radiator. Interference effects for
radiation from transitions in periodic arrangements  cause an effective
threshold behaviour at a value of $\gamma \approx 1000$. These effects
also produce a frequency dependent photon yield. The foil thickness must
be comparable to or larger than the formation zone
\beq
D = \gamma c/\omega_p
\eeq
which in practical situations ($\hbar \omega_p = 20\,$eV; $\gamma = 5
\cdot 10^3$) is about 50\,$\mu$m.
Transition radiation detectors are mainly used for
$e/\pi$-setaration. In cosmic ray experiments transition radiation
emission can also be employed to measure the energy of muons in the
TeV-range. 

%
\subsection*{Bremsstrahlung}
If a charged particle is decelerated in the Coulomb field of a nucleus a
fraction of its kinetic energy will be emitted in form of real photons
(bremsstrahlung). The energy loss by bremsstrahlung for high energies 
can be described by \cite{r:cgru:[2]}
\beq
-\frac{{\rm d}E}{{\rm d}x} = 4 \alpha N_A \frac{Z^2}{A} \cdot z^2 r^2 E \ln
\frac{183}{Z^{1/3}} \quad,
\label{E:cgru:65}
\eeq
where $r= \frac{1}{4\pi \varepsilon_0} \cdot \frac{e^2}{m c^2}$.
Bremsstrahlung is mainly produced by electrons because
\beq
r_e \propto \frac{1}{m_e} \quad.
\eeq
Equation \ref{E:cgru:65} can be rewritten for electrons
\beq
-\frac{{\rm d}E}{{\rm d}x} = \frac{E}{X_0} \quad,
\eeq
where
\beq
X_0 = \frac{A}{4 \alpha N_A Z (Z+1) r_e^2 \ln (183\, Z^{-1/3})}
\label{E:cgru:68}
\eeq
is the radiation length of the absorber in which bremsstrahlung is
produced. Here we have included also radiation from electrons  ($\sim Z$,
because there are $Z$ electrons per nucleus). If screening effects are
taken into account $X_0$ can be more accurately described by 
\cite{r:cgru:[6]}
\beq
X_0 = \frac{716.4\,A}{Z (Z+1) \ln (287/\sqrt{Z})} \quad [{\rm g/cm}^2] \quad.
\eeq
The important point about bremsstrahlung is that the energy loss is
proportional to the energy. The energy where the losses due to ionization
and bremsstrahlung for electrons are the same is called critical energy
\beq
\left. \frac{{\rm d}E_c}{{\rm d}x} \right|_{\rm ion} = \left. \frac{{\rm
d}E_c}{{\rm d}x} \right|_{\rm brems} \quad.
\eeq
For solid or liquid absorbers the critical energy can be approximated by 
\cite{r:cgru:[6]}
\beq
E_c = \frac{610\,{\rm MeV}}{Z + 1.24} \quad,
\eeq
while for gases one has \cite{r:cgru:[6]}
\beq
E_c = \frac{710\,{\rm MeV}}{Z + 0.92} \quad.
\eeq
The difference between gases on the one hand and solids and liquids on
the other hand comes about because the density corrections are different
in these substances, and this modifies $\left. \frac{{\rm d}E}{{\rm d}x}
\right|_{\rm ion}$.

The energy spectrum of bremsstrahlung photons is $\sim E_{\gamma}^{-1}$,
where $E_{\gamma}$ is the photon energy.

At high energies also radiation
from heavier particles becomes important  and consequently a critical energy
for these particles can be defined. Since
\beq
\left. \frac{{\rm d}E}{{\rm d}x} \right|_{\rm brems}  \propto
\frac{1}{{\rm m}^2}
\eeq
the critical energy e.g.\ for muons in iron is
\beq
E_c = \frac{610\,{\rm MeV}}{Z + 1.24} \cdot \left( \frac{m_{\mu}}{m_e}
\right)^2 = 960\,{\rm GeV} \quad.
\eeq

\subsection*{Direct Electron Pair Production}
Direct electron pair production in  the Coulomb field of a nucleus
via virtual photons (``tridents'')  is a dominant energy loss mechanism at
high energies. The energy loss for singly charged particles due to this
process
can be represented by

\beq
\left. -\frac{{\rm d}E}{{\rm d}x} \right|_{\rm pair}  = b (Z, A, E) \cdot E 
\quad.
\eeq

It is essentially - like bremsstrahlung - also proportional to the 
particle's energy. Because
bremsstrahlung and direct pair production dominate at high energies this
offers an attractive possibility to build also muon calorimeters 
\cite{r:cgru:[2]}.
The average rate of muon energy losses can be parametrized as

\beq
\frac{{\rm d}E}{{\rm d}x} = a(E) + b(E) \cdot E
\eeq
where $a(E)$ represents the ionization energy loss and $b(E)$ is the sum of
direct elektron pair production, bremsstrahlung and photonuclear 
interactions. 

\begin{figure}[ht!] 
\begin{center}
\includegraphics[width=0.9\textwidth,clip]
{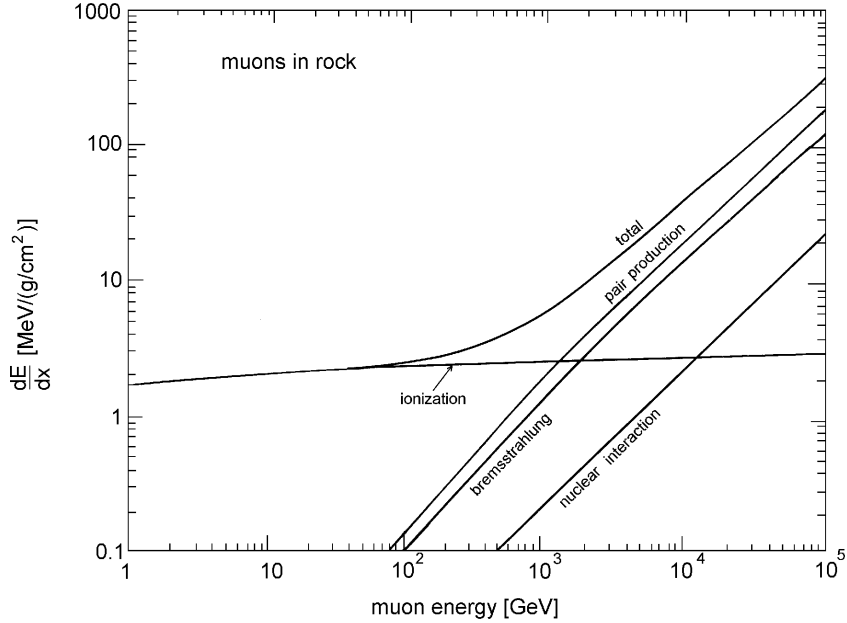}
\caption{Contributions to the energy loss of muons in standard rock (Z
  = 11; A = 22; $\rho = 3g/cm^3$).}
\label{F:cgru:new10}
\end{center}
\end{figure}

The various contributions to the energy loss of muons in standard rock
(Z = 11; A = 22; $\rho = 3g/cm^3$) are shown in figure
\ref{F:cgru:new10}. 

%
\subsection*{Nuclear Interactions}
Nuclear interactions play an important role in the detection of  neutral
particles other than photons. They are also responsible for the
development of hadronic cascades. The total cross section for nucleons is
of the order of 50\,mbarn and varies slightly with energy. It
has an elastic ($\sigma_{\rm el}$) and inelastic part ($\sigma_{\rm inel}$).
The inelastic cross section has a material dependence
\beq
\sigma_{\rm inel} \approx \sigma_0 A^{\alpha}
\eeq
with $\alpha = 0.71$. The corresponding absorption length $\lambda_a$ is
\cite{r:cgru:[2]}
\beq
\lambda_a = \frac{A}{N_A \cdot \rho \cdot \sigma_{\rm inel}} [{\rm cm}]
\eeq
($A$ in g/mol, $N_A$ in mol$^{-1}$, $\rho$ in g/cm$^3$, and $\sigma_{\rm
inel}$ in cm$^2$). \\
This quantity has to be distinguished from the nuclear interaction
length $\lambda_w$, which is related to the total cross section
\beq
\lambda_w = \frac{A}{N_A \cdot \rho \cdot \sigma_{\rm total}} [{\rm cm}]
\quad.
\eeq
Since $\sigma_{\rm total} > \sigma_{\rm inel}$, $\lambda_w <
\lambda_a$ holds.

Strong interactions have a multiplicity which grows  logarithmically with
energy. The particles  are produced  in a narrow cone around the forward
direction with an average transverse momentum of $p_T = 350\,$MeV/c,
which is responsible for the lateral spread of hadronic cascades.

A useful relation for the calculation of interaction rates per (g/cm$^2$) is
\beq
\phi (({\rm g/cm}^2)^{-1}) = \sigma_N \cdot N_A
\eeq
where $\sigma_N$ is the cross section per nucleon and $N_A$
Avogadro's number.

\section*{Interaction of Photons}
Photons are attenuated in matter via the processes of the photoelectric
effect, Compton scattering and pair production. The intensity of a photon
beam varies in matter according to
\beq
I = I_0 \:{\rm e}^{-\mu x} \quad,
\eeq
where $\mu$ is mass attenuation coefficient. $\mu$ is related to the
photon cross sections $\sigma_i$ by
\beq
\mu = \frac{N_A}{A} \sum^3_{i=1} \sigma_i \quad.
\eeq

\subsection*{Photoelectric Effect}
Atomic electrons can absorb the energy of a photon completely
\beq
\gamma + \mbox{atom} \rightarrow \mbox{atom}^+ + e^- \quad.
\eeq
The cross section for absorption of a photon of energy $E_{\gamma}$ is
particularly large in the $K$-shell (80\% of the total cross section). The
total cross section for photon absorption in the $K$-shell is
\beq
\sigma^K_{\rm Photo} = \left( \frac{32}{\varepsilon^7} \right)^{1/2} \alpha^4
Z^5 \sigma_{\rm Thomson} [{\rm cm}^2/{\rm atom}] \quad,
\eeq
where $\varepsilon = E_{\gamma}/m_ec^2$, and $\sigma_{\rm Thomson} =
\frac{8}{3} \pi r_e^2 = 665$\,mbarn is the cross section for Thomson
scattering. For high energies the energy dependence becomes softer
\beq
\sigma^K_{\rm Photo} = 4 \pi r_e^2 Z^5 \alpha^4 \cdot \frac{1}{\varepsilon}
\quad.
\eeq
The photoelectric cross section has sharp discontinuities when
$E_{\gamma}$ coincides with the binding energy of atomic shells. As a
consequence of a photoabsorption in the $K$-shell characteristic X-rays
or Auger electrons are emitted \cite{r:cgru:[2]}.

\subsection*{Compton Scattering}
The Compton effect describes the scattering of photons off  quasi-free
atomic electrons
\beq
\gamma + e \rightarrow \gamma' + e' \quad.
\eeq
The cross section for this process, given by the Klein-Nishina formula, can
be approximated at high energies by
\beq
\sigma_c \propto \frac{\ln \varepsilon}{\varepsilon} \cdot Z
\eeq
where $Z$ is the number of electrons in the target atom. From energy and
momentum conservation one can derive the ratio of scattered
($E_{\gamma}'$) to incident photon energy ($E_{\gamma}$)
\beq
\frac{E_{\gamma}'}{E_{\gamma}} = \frac{1}{1+\varepsilon(1-\cos
\Theta_{\gamma})} \quad,
\eeq
where $\Theta_{\gamma}$ is the scattering angle of the photon with
respect to its original direction.

For backscattering ($\Theta_{\gamma} = \pi$) the energy transfer to the
electron $E_{\rm kin}$ reaches a maximum value
\beq
E^{\max}_{\rm kin} = \frac{2 \varepsilon^2}{1+2 \varepsilon} m_ec^2 \quad,
\eeq
which, in the extreme case ($\varepsilon \gg 1$), equals $E_{\gamma}$.

In Compton scattering only a fraction of the photon energy is transferred to
the electron. Therefore, one defines an energy scattering cross section
\beq
\sigma_{cs} =  \frac{E_{\gamma}'}{E_{\gamma}} \sigma_c
\eeq
and an energy absorption cross section
\beq
\sigma_{ca} = \sigma_c - \sigma_{cs} = \sigma_c \frac{E_{\rm
kin}}{E_{\gamma}} \quad.
\eeq
At accelerators and in astrophysics also the process of inverse   Compton
scattering is of importance \protect\cite{r:cgru:[2]}.

\subsection*{Pair Production}
The production of an electron-positron pair in the Coulomb field of a
nucleus requires a certain minimum energy
\beq
E_{\gamma} \geq 2 m_ec^2 + \frac{2 m_e^2c^2}{m_{\rm nucleus}} \quad.
\eeq
Since for all practical cases $m_{\rm nucleus} \gg m_e$, one has effectively
$E_{\gamma} \geq 2 m_e c^2$.

The total cross section in the case of complete screening $\left( \varepsilon
\gg \frac{1}{\alpha Z^{1/3}} \right)$; i.e.\ at reasonably high energies
$(E_{\gamma} \gg 20\,$MeV), is
\beq
\sigma_{\rm pair} = 4 \alpha r^2_e Z^2 \left( \frac{7}{9} \ln
\frac{183}{Z^{1/3}} - \frac{1}{54} \right)\quad [{\rm cm}^2/{\rm atom}] \quad.
\label{E:cgru:104}
\eeq
Neglecting the small additive term $1/54$ in eq. \ref{E:cgru:104} one can 
rewrite,
using eq. \ref{E:cgru:65} and eq. \ref{E:cgru:68},
\beq
\sigma_{\rm pair} = \frac{7}{9} \frac{A}{N_A} \cdot \frac{1}{X_0} \quad.
\eeq
The partition of the energy to the electron and positron is symmetric at low
energies $(E_{\gamma} \ll 50\,$MeV) and increasingly asymmetric at high
energies $(E_{\gamma} > 1\,$GeV) \protect\cite{r:cgru:[2]}.

\begin{figure}[ht!]
  \begin{center}
 \includegraphics[angle=90,width=0.6\textwidth]{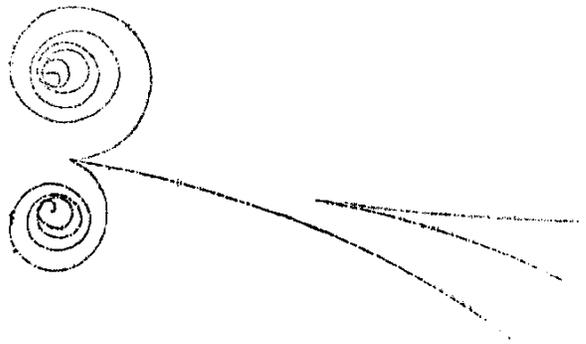}
\vspace*{0.5cm}
    \caption{Photoproduction in the Coulomb-field of an electron 
($\gamma + e^- \rightarrow e^+ + e^- + e^-$) and on a nucleus 
($ \gamma + \mbox{nucleus} \rightarrow e^+ + e^- +
\mbox{nucleus}^\prime$)
\protect\cite{r:cgru:Close}}
    \label{F:cgru:pairprod}
  \end{center}
\end{figure}

Figure \ref{F:cgru:pairprod} shows the photoproduction  of an
electron-positron pair in the Coulomb-field of an electron ($\gamma +
e^- \rightarrow e^+ + e^- + e^-$) and also a pair-production in the
field of a nucleus ($ \gamma + \mbox{nucleus} \rightarrow e^+ + e^- +
\mbox{nucleus}^\prime$) \cite{r:cgru:Close}.

\newpage
\subsection*{Mass-Attenuation Coefficients}
\begin{figure}[bh!]
\begin{center}
\begin{minipage}[b]{.45\textwidth}
  \begin{center}
    \leavevmode
\includegraphics[angle=0.2,totalheight=7.0cm,width=\textwidth]{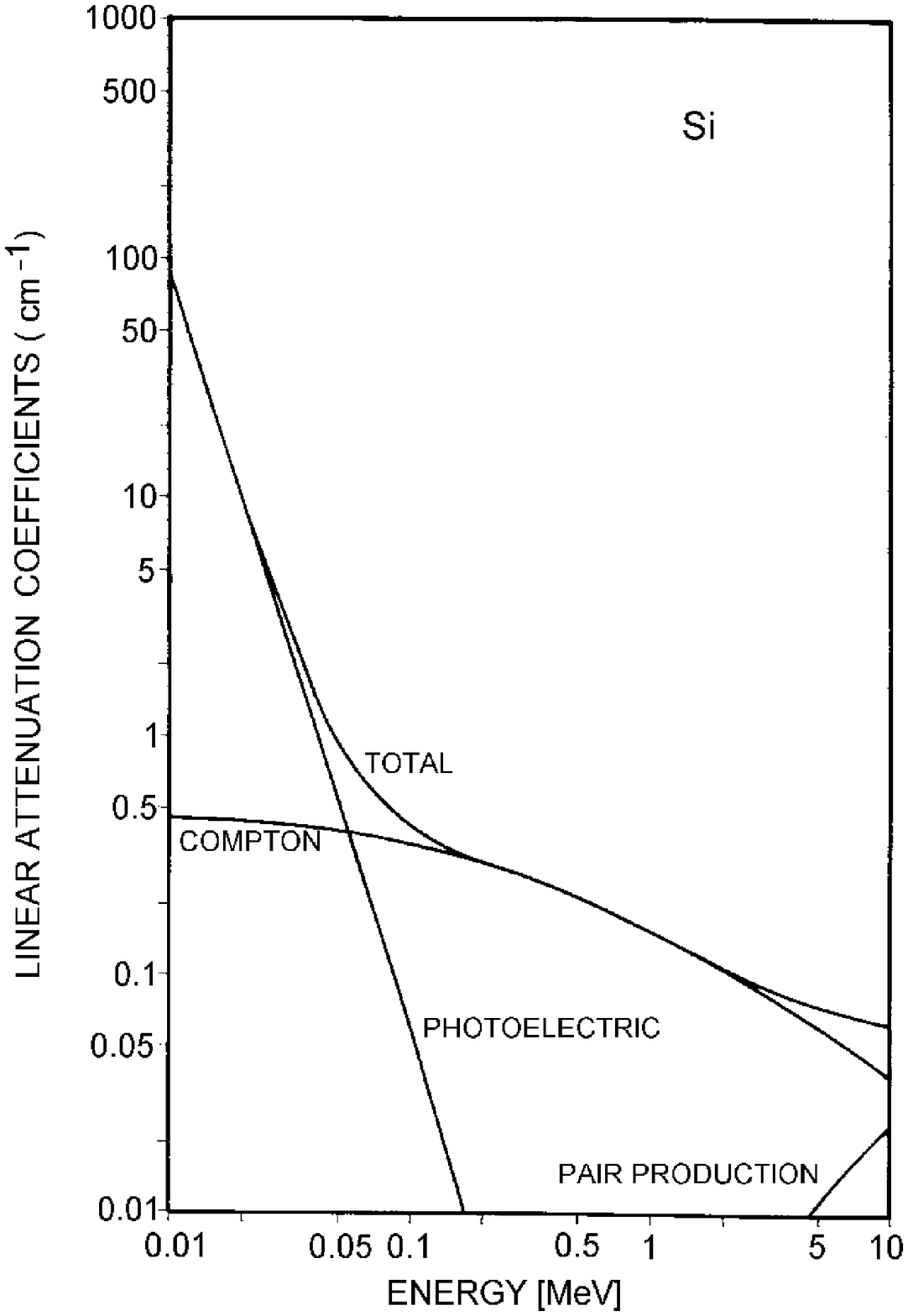}
\vspace{6pt}
\caption{Mass attenuation coefficients for photon interactions in silicon 
\protect\cite{r:cgru:[30]}}
\label{F:cgru:34}
  \end{center}
\end{minipage}
\hspace*{.04\textwidth}
\begin{minipage}[b]{.45\textwidth}
  \begin{center}
    \leavevmode
\includegraphics[totalheight=7.0cm,width=\textwidth,clip]
{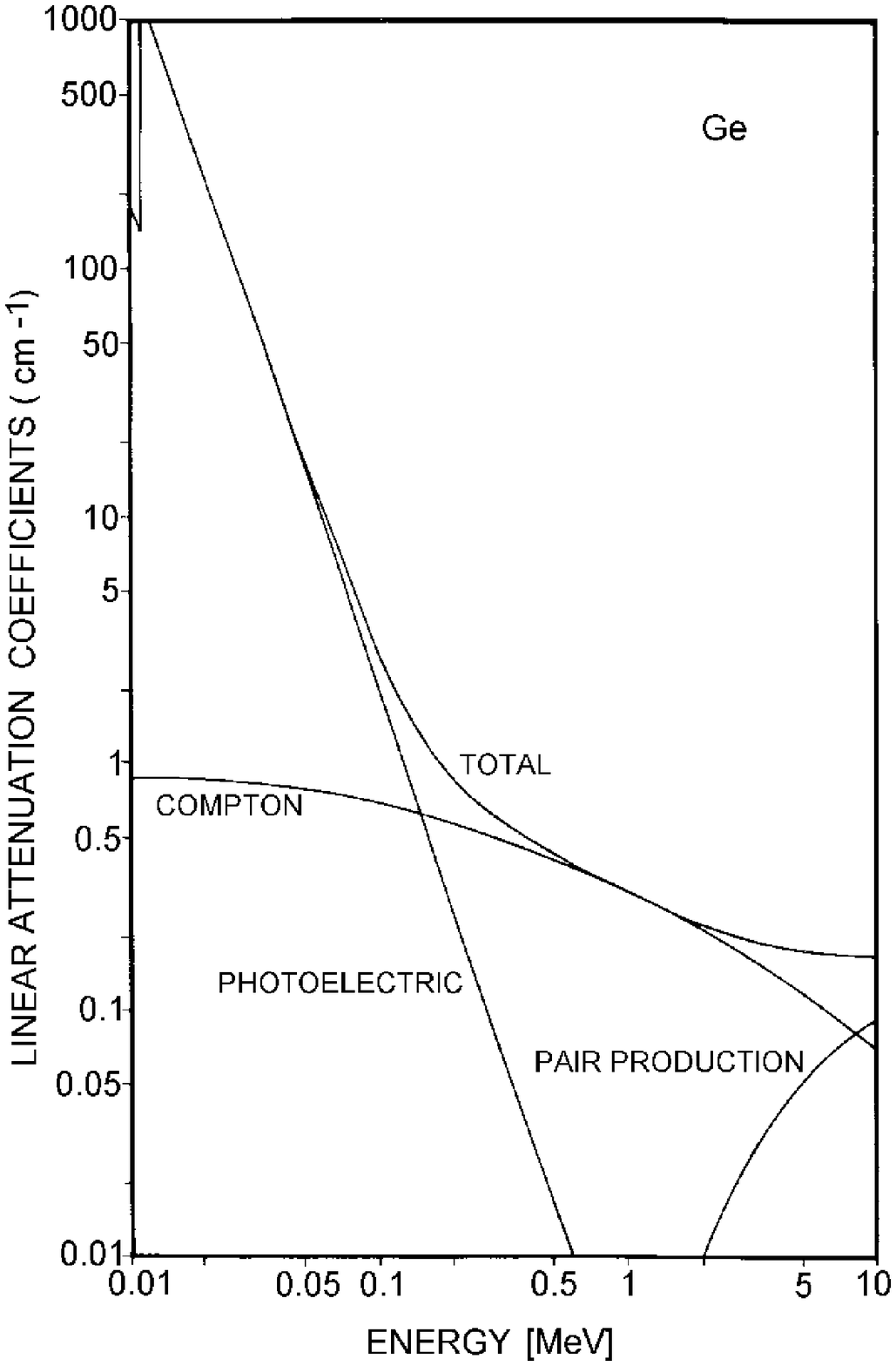}
\vspace{6pt}
\caption{Mass attenuation coefficients for photon interactions in
germanium \protect\cite{r:cgru:[30]}}
\label{F:cgru:35}
  \end{center}
\end{minipage}
\vspace*{1.5cm}
\begin{minipage}[b]{.45\textwidth}
  \begin{center}
    \leavevmode
\includegraphics[angle=0.2,totalheight=7.0cm,width=\textwidth,clip]
{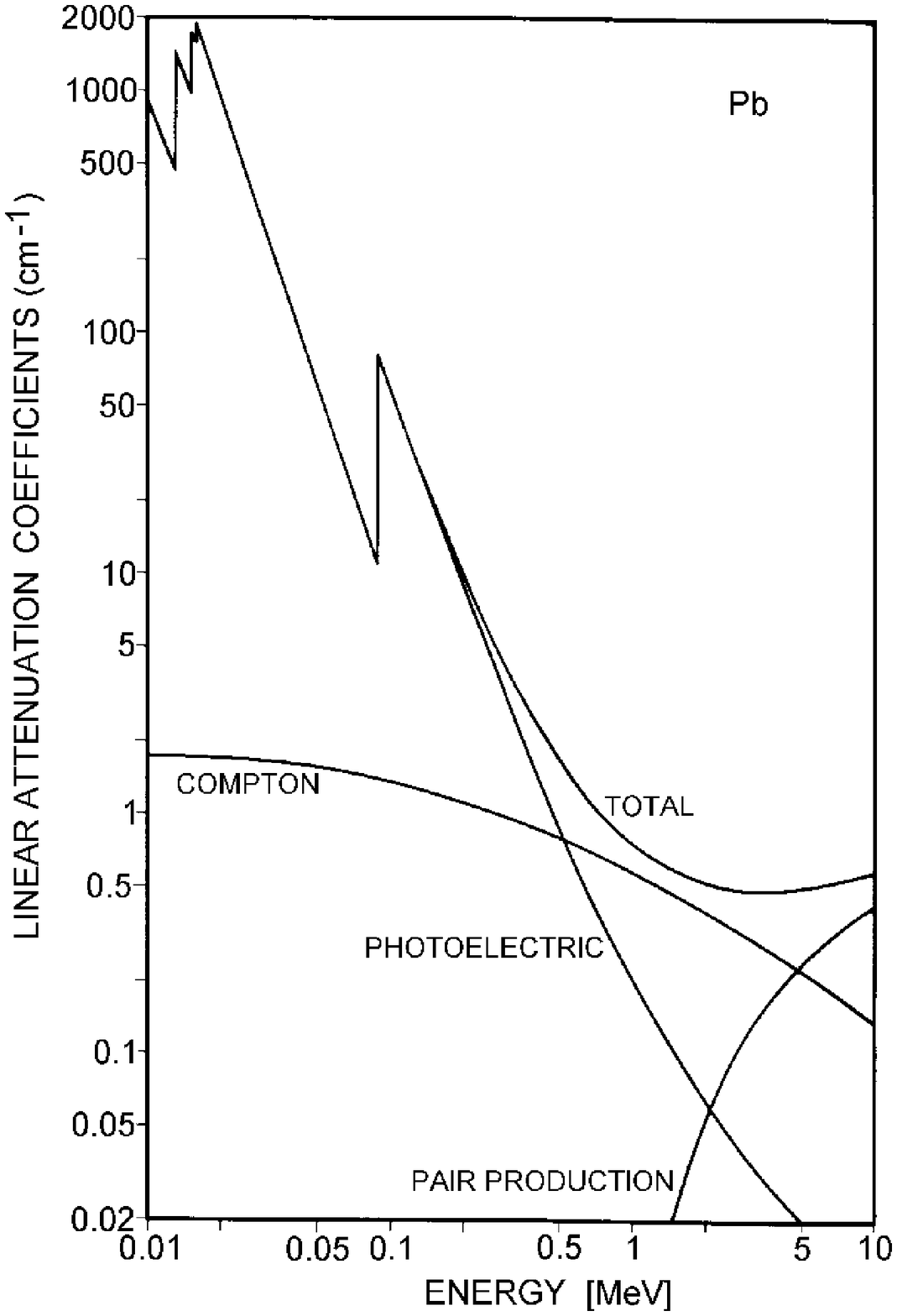}
\vspace{6pt}
\caption{Mass attenuation coefficients for photon interactions in lead 
\protect\cite{r:cgru:[30]}}
\label{F:cgru:36}
  \end{center}
\end{minipage}
\hspace*{.04\textwidth}
\begin{minipage}[b]{.45\textwidth}
The mass-attenuation coefficients for photon interactions are shown in
figures \ref{F:cgru:34}-\ref{F:cgru:36} for silicon, 
germanium and lead \cite{r:cgru:[30]}. The photoelectric effect
dominates at low energies ($E_{\gamma} < 100\,$keV). Superimposed on the 
continuous photoelectric attenuation coefficient are absorption edges
characteristic of the absorber material. Pair production
dominates at high energies ($>10$\,MeV). In the intermediate region
Compton scattering prevails.

    \vspace{1.1cm}
\end{minipage}
\end{center}
\end{figure}

\section*{Interaction of Neutrons}
In the same way as photons are detected via their interactions also
neutrons have to be measured indirectly. Depending on the neutron energy
various reactions can be considered which produce charged particles
which are then detected via their ionization or scintillation 
\cite{r:cgru:[2]}.
\begin{itemize}
\item[a)]
Low energies ($< 20\,$MeV)
\beqar
n + {^6{\rm Li}} & \rightarrow & \alpha + {^3{\rm H}} \nonumber \\
n + {^{10}{\rm B}} & \rightarrow & \alpha + {^7{\rm Li}} \nonumber \\
n + {^3{\rm He}} & \rightarrow & p + {^3{\rm H}}  \\
n + p & \rightarrow & n+p \nonumber
\eeqar
The conversion material can be a component of a scintillator (e.g.\ LiI (Tl)),
a
thin layer of material in front of the sensitive volume of a gaseous detector
(boron layer), or an admixture to the counting gas of a proportional counter
(BF$_3$,  $^3$He, or protons in CH$_4$).
\item[b)]
Medium energies ($20\,{\rm MeV} \leq E_{\rm kin} \leq 1\,$GeV) \\
The ($n,p$)-recoil reaction can be used for neutron detection in detectors
which contain many quasi-free protons in their sensitive volume (e.g.\
hydrocarbons).
\item[c)]
High energies ($E > 1\,$GeV) \\
Neutrons of high energy initiate hadron cascades in inelastic interactions
which are easy to identify in hadron calorimeters.
\end{itemize}

Neutrons are detected with relatively high efficiency at very low energies.
Therefore, it is often useful to slow down neutrons with substances
containing many protons, because neutrons can transfer a large amount of
energy to collision partners of the same mass. 
In some fields of application, like in radiation 
protection at
nuclear reactors, it is of importance to know the energy of fission
neutrons, because  the relative  biological effectiveness depends on it.
This can e.g.\ be achieved with a stack of plastic  detectors interleaved with
foils of materials with different threshold energies for neutron conversion
\cite{r:cgru:[28]}.

\section*{Interactions of Neutrinos}
Neutrinos are very difficult to detect. Depending on the neutrino flavor the
following inverse beta decay like interactions can be considered:
\beqar
\nu_e + n & \rightarrow & p + e^- \nonumber \\
\bar{\nu}_e + p & \rightarrow & n + e^+ \vspace{1mm}\nonumber \\
\nu_{\mu} + n & \rightarrow & p + \mu^- \nonumber \\
\bar{\nu}_{\mu} + p & \rightarrow & n + \mu^+ \vspace{1mm}\\
\nu_{\tau} + n & \rightarrow & p + \tau^- \nonumber \\ 
\bar{\nu}_{\tau} + p & \rightarrow & n + \tau^+ \nonumber
\eeqar
The cross section for $\nu_e$-detection in the MeV-range can be
estimated as \protect\cite{r:cgru:[32]}
\beqar
\sigma (\nu_eN) & = & \frac{4}{\pi} \cdot 10^{-10} \left( \frac{\hbar p}{(m_p
c)^2} \right)^2  \nonumber \\
& = & 6.4 \cdot 10^{-44} {\rm cm}^2 \mbox{ for 1\,MeV} \quad.
\eeqar
This means that the interaction probability of e.g.\ solar neutrinos in a water
Cherenkov counter of $d=100$\,meter  thickness is only
\beq
\phi = \sigma \cdot N_A \cdot d = 3.8 \cdot 10^{-16} \quad.
\eeq
Since the coupling constant of weak interactions has a dimension of
1/GeV$^2$, the neutrino cross section must rise at high energies like the
square of the center-of-mass energy. For fixed target experiments we can
parametrize
\beqar
\sigma (\nu_{\mu} N) & = & 0.67 \cdot 10^{-38} E_{\nu} [{\rm GeV}] \quad {\rm
cm}^2/{\rm nucleon} \nonumber \\
\sigma (\bar{\nu}_{\mu} N) & = & 0.34 \cdot 10^{-38} E_{\nu} [{\rm GeV}] \quad
{\rm cm}^2/{\rm nucleon}
\eeqar
This shows that even at 100 GeV the neutrino cross section is lower by 11
orders of magnitude compared to the total proton-proton cross section.

\section*{Electromagnetic Cascades}
The development of cascades induced by electrons, positrons or photons
is governed by bremsstrahlung of electrons and pair production of
photons. Secondary particle production continues until photons fall below
the pair production threshold, and energy losses of electrons other than
bremsstrahlung start to dominate: the number of shower particles decays
exponentially.

Already a very simple model can describe the main features
of particle multiplication in electromagnetic cascades: A photon
of energy $E_0$ starts the cascade by producing an $e^+e^-$-pair after
one radiation length. Assuming that the energy is shared symmetrically
between the particles at each multiplication step, one gets at the depth $t$
\beq
N(t) = 2^t
\eeq
particles with energy
\beq
E(t) = E_0 \cdot 2^{-t} \quad.
\eeq
The multiplication continues until the electrons fall below the critical
energy $E_c$
\beq
E_c = E_0 \cdot 2^{-t_{\max}} \quad.
\label{E:cgru:112}
\eeq
From then on ($t > t_{\max}$) the shower particles are only absorbed. The
position of the shower maximum is obtained from eq. \ref{E:cgru:112}
\beq
t_{\max} = \frac{\ln E_0/E_c}{\ln 2} \propto \ln E_0 \quad.
\eeq
The total number of shower particles is
\beqar
S & = & \sum^{t_{\max}}_{t=0} N(t) = \sum 2^t = 2^{t_{\max}+1} - 1 \approx
2^{t_{\max}+1} \nonumber \\
S & = & 2 \cdot 2^{t_{\max}} = 2 \cdot \frac{E_0}{E_c} \propto E_0 \quad.
\eeqar
If the shower particles are sampled in steps $t$ measured in  units of
$X_0$, the total track length is obtained as
\beq
S^{\ast} = \frac{S}{t} = 2 \frac{E_0}{E_c} \cdot \frac{1}{t} \quad,
\eeq
which leads to an energy resolution of
\beq
\frac{\sigma}{E_0} = \frac{\sqrt{S^{\ast}}}{S^{\ast}} = \frac{\sqrt{t}}{\sqrt{2
E_0/E_c}} \propto \frac{\sqrt{t}}{\sqrt{E_0}} \quad.
\eeq
In a more realistic description the longitudinal development of the electron
shower can be approximated by \cite{r:cgru:[6]}
\beq
\frac{{\rm d}E}{{\rm d}t} = \mbox{const} \cdot t^a  \cdot {\rm e}^{-bt} \quad,
\eeq
where $a$, $b$ are fit parameters.

\begin{figure}[ht!]
  \begin{center}
    \includegraphics[width=0.7\textwidth
     ,clip]
{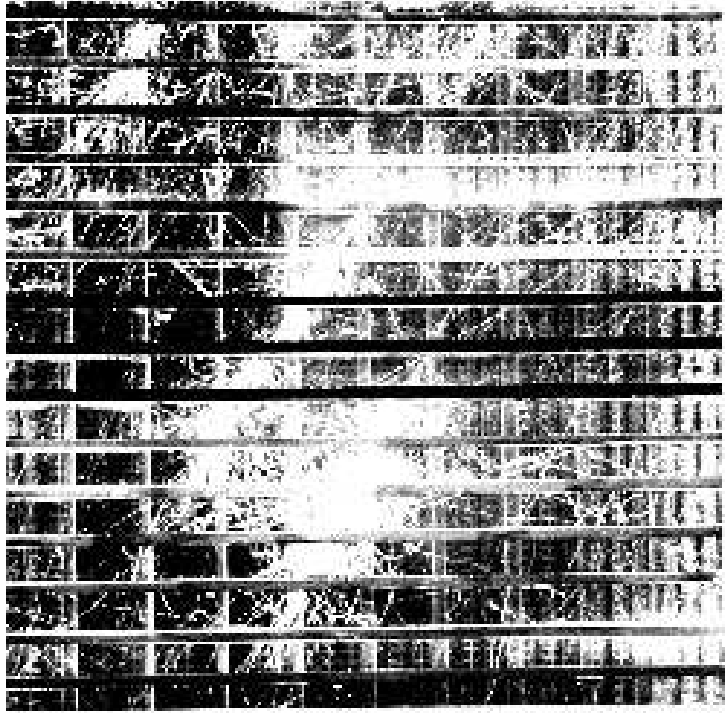}
\vspace*{0.5cm}
    \caption{Some muon induced electromagnetic cascades in a
      multi-plate cloud chamber operated in a concrete shielded air
      shower array \protect\cite{r:cgru:[new34]}}
    \label{F:cgru:new14}
  \end{center}
\end{figure}

Figure \ref{F:cgru:new14} shows muon induced electromagnetic 
cascades in a multi-plate
cloud chamber \cite{r:cgru:[new34]}.

The lateral spread of an electromagnetic shower is mainly caused by
multiple scattering. It is described by the Moli\`{e}re radius
\beq
R_m = \frac{21\,{\rm MeV}}{E_c} X_0 \: [{\rm g/cm}^2] \quad.
\eeq
95\% of the shower energy in a homogeneous calorimeter is contained in a
cylinder of radius $ 2 R_m$ around the shower axis.

\begin{figure}[!ht]
\begin{center}
    \includegraphics[width=0.7\textwidth,angle=0.0
     ,clip]{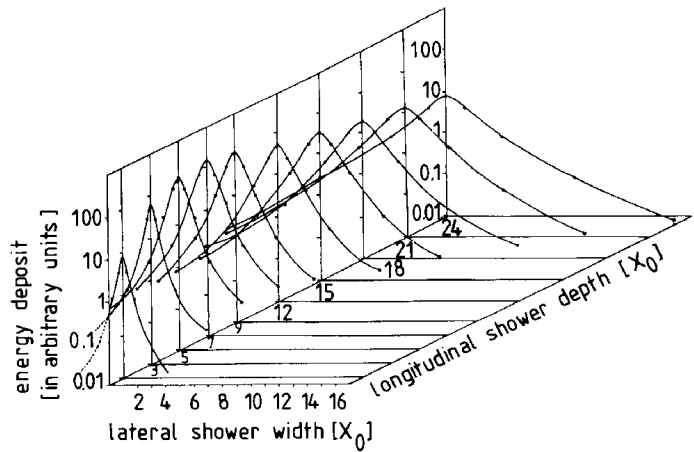}
    \vspace*{0.2cm}
    \caption{Sketch of the longitudinal  and lateral development of an
    electromagnetic cascade in a homogeneous absorber 
    \protect\cite{r:cgru:[2]}}
    \label{F:cgru:45}
\end{center}
\end{figure}

Figure \ref{F:cgru:45} demonstrates the interplay of the longitudinal
and lateral
development of an electromagnetic shower \cite{r:cgru:[2]}.

\newpage
\section*{Hadron Cascades}
The longitudinal development of electromagnetic cascades is
characterized by the radiation length $X_0$ and their lateral width is
determined by multiple scattering. In contrast to this, hadron showers are
governed in their longitudinal structure by the nuclear interaction length
$\lambda$ and by transverse momenta of secondary particles as far as
lateral width is
concerned. Since for most materials $\lambda \gg X_0$, and $\langle
p_T^{\rm interaction}\rangle \gg \langle p_T^{\rm multiple\:
scattering}\rangle$ hadron showers are longer and wider.

Part of the energy of the incident hadron is spent to break up nuclear
bonds. This fraction of the energy is invisible in hadron calorimeters.
Further energy is lost by escaping particles like neutrinos and muons as a
result of hadron decays. Since the fraction of lost binding energy and
escaping particles fluctuates considerably, the energy resolution of hadron
calorimeters is systematically inferior to electron calorimeters.

\begin{figure}[ht!]
  \begin{center}
    \includegraphics[angle=0.5
,width=.5\textwidth]
          {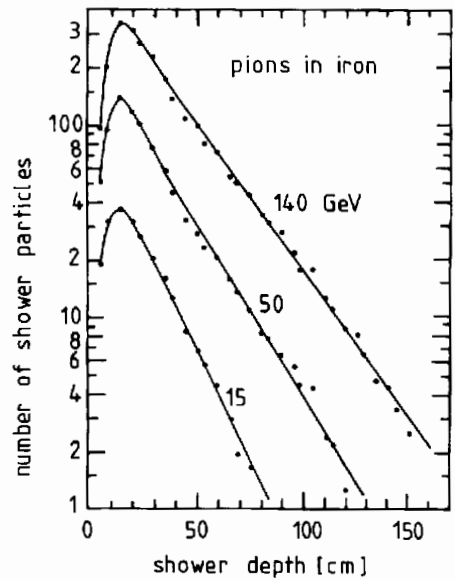}
    \vspace{5pt}
    \caption{Longitudinal development of pion induced hadron cascades 
      \protect\cite{r:cgru:[37]}}
    \label{F:cgru:new16}
  \end{center}
\end{figure}

\begin{figure}[ht!]
  \begin{center}
    \includegraphics[width=0.7\textwidth,clip]
{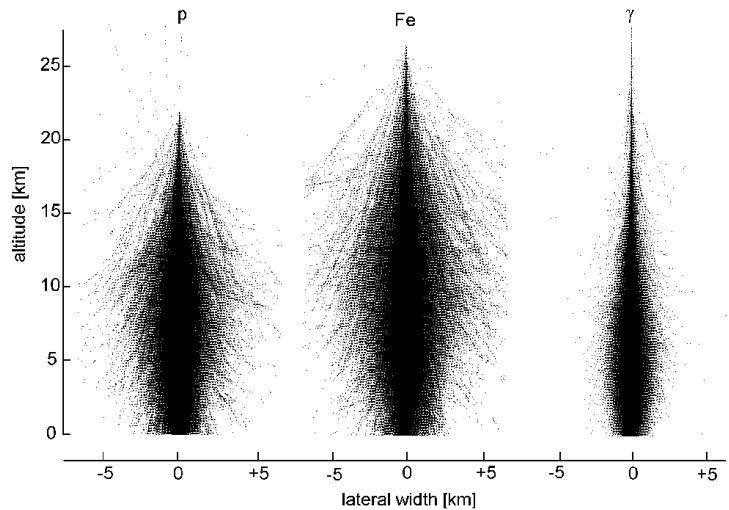}
\vspace*{0.3cm}
    \caption{Comparison between proton, iron, and photon induced
      cascades in the atmosphere. The primary energy in each case is
      10$^{14}$eV \protect\cite{r:cgru:knapp}.}
    \label{F:cgru:new17}
  \end{center}
\end{figure}

The longitudinal development of pion induced hadron cascades is plotted in
figure \ref{F:cgru:new16}.
Figure \ref{F:cgru:new17} 
shows a comparison between proton, iron, and
photon induced cascades in the atmosphere \cite{r:cgru:knapp}.

The different response of calorimeters to electrons and hadrons is an
undesirable feature for the energy measurement of jets of unknown
particle composition. By appropriate compensation techniques, however,
the electron to hadron response can be equalized.

\section*{Particle Identification}
\begin{figure}[ht!]
  \begin{center}
     \includegraphics[bb=81 272 521 562, 
width=.7\textwidth,clip]{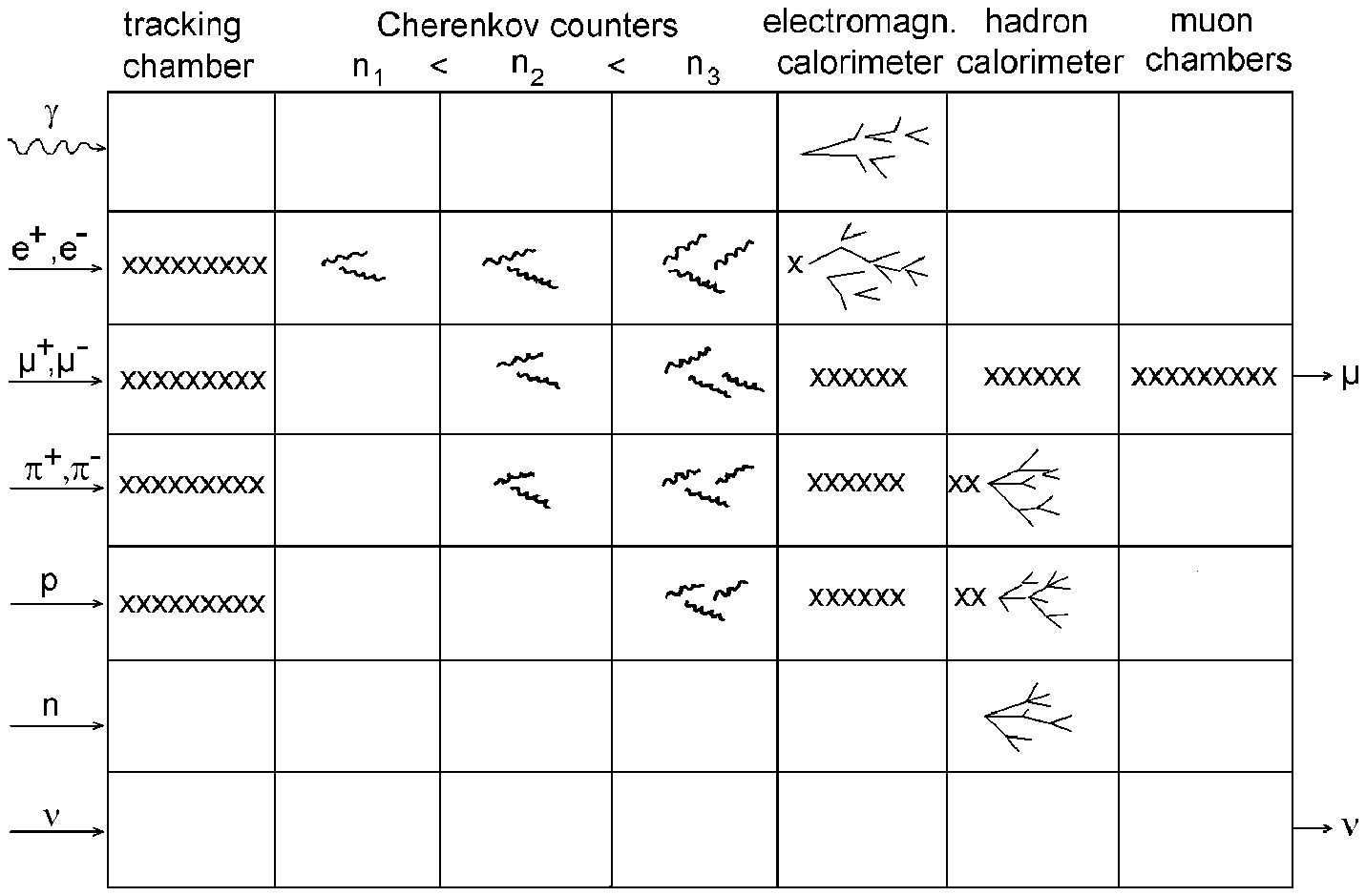}
    \caption{Particle identification using a detector consisting of a
      tracking chamber, Cherenkov counters, calorimetry and muon chambers.}
    \label{F:cgru:new18}
  \end{center}
\end{figure}

Particle identification is based on measurements which are sensitive
to the particle velocity, its charge and its momentum.
Figure \ref{F:cgru:new18} sketches the different possibilities to
separate photons, electrons, positrons, muons, charged pions, protons, 
neutrons and neutrinos in a mixed particle beam using a general
purpose detector.

\begin{figure}[ht!]
  \begin{center}
\vspace*{-0.5cm}
\includegraphics[width=0.55\textwidth,clip]
    {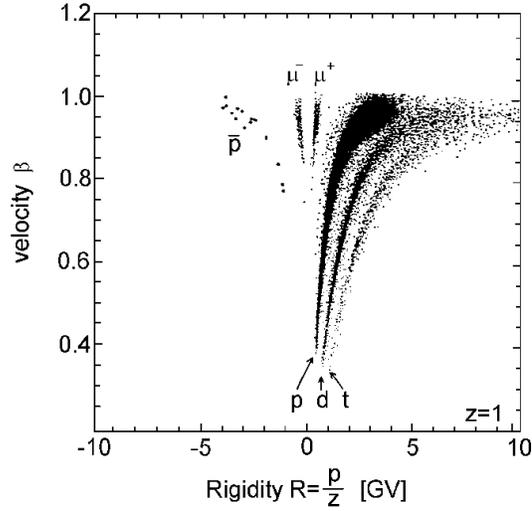}
    \caption{Particle identification in a balloon borne experiment using
      momentum, time-of-flight, dE/dx and Cherenkov radiation
      information \protect\cite{r:cgru:Mitchell}.}
    \label{F:cgru:new19}
  \end{center}
\end{figure}

Figure \ref{F:cgru:new19} shows the particle separation power of a
balloon borne experiment using momentum, time-of-flight, dE/dx and
Cherenkov radiation measurements \cite{r:cgru:Mitchell}.

\begin{figure}[ht!]
  \begin{center}
    \includegraphics[
width=0.8\textwidth,
clip]
{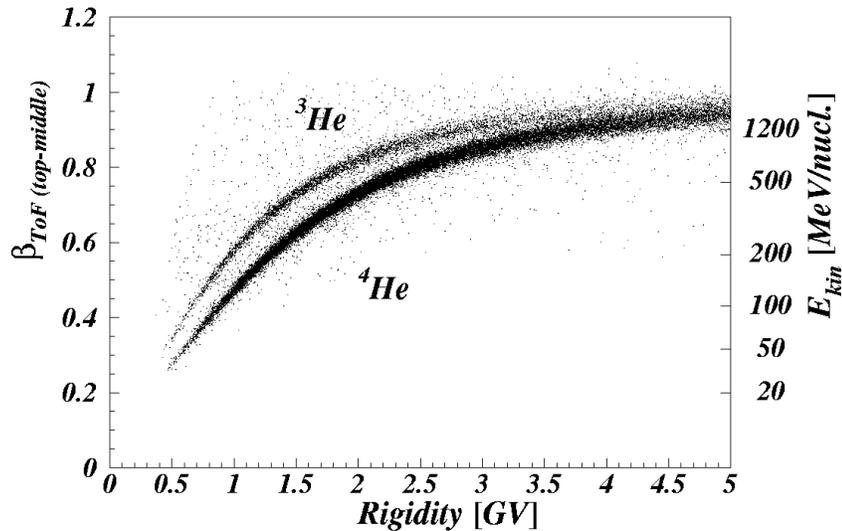}
    \caption{Isotopic abundance of energetic cosmic ray helium nuclei
      \protect\cite{r:cgru:Simon}}
    \label{F:cgru:new20}
  \end{center}
\end{figure}

Even the abundance of different helium isotopes can be determined from 
a velocity and momentum measurement (figure \ref{F:cgru:new20}
\cite{r:cgru:Simon}).
This is feasable, because at fixed momentum the lighter isotope $^3$He 
is faster than the more abundant $^4$He.

\clearpage
\section*{Conclusion}
Basic physical principles can be used to identify all kinds of  elementary
particles and nuclei. The precise measurement of the particle composition
in high energy physics experiments at accelerators and in cosmic rays is
essential for the insight into the underlying physics processes. This is an
important ingredient for the progress in the fields of elementary particles
and astrophysics aiming at the unification of forces and the understanding
of the evolution of the universe.

\section*{Acknowledgements}
The author thanks 
Mrs.\ C.\ Hauke (figures) and
Dipl.\ Phys.\ G.\ 
Prange (text and layout) for their help in preparing the manuscript.

\end{document}